\documentclass[12pt]{article}
\usepackage{epsfig}
\usepackage{graphicx}

\usepackage{times}

\newenvironment{sciabstract}{%
\begin{quote} \bf}
{\end{quote}}

\newcounter{lastnote}

\title{Octave-spanning semiconductor laser}

\author
{M.R\"osch$^{1}$ .G. Scalari$^{1,\ast}$, M. Beck $^1$, J. Faist$^1$\\
\\
\normalsize{$^1$ Institute of Quantum Electronics, Eidgen\"ossische  Technische Hochschule Z\"urich, Switzerland}\\
\\
\normalsize{$^\ast$E-mail:  scalari@phys.ethz.ch}
}
\date{}

\begin{document} 


\baselineskip24pt


\maketitle


\begin{sciabstract}
We present here a semiconductor injection laser operating in  continuous wave with an emission covering more than one octave in frequency, and displaying homogeneous power distribution among the lasing modes. The gain medium is based on a heterogeneous quantum cascade structure  operating in the THz range. Laser emission in continuous wave takes place  from 1.64 THz to 3.35 THz with optical powers in the mW range and more than 80 modes above threshold. Free-running beatnote investigations on narrow waveguides with linewidths of 980 Hz limited by jitter indicate frequency comb operation on a spectral bandwidth as wide as 624 GHz, making such devices ideal candidates for octave-spanning semiconductor-laser-based THz frequency combs.
\end{sciabstract}

%
%
%
%
%
%
A broadband gain medium is desirable for a wide range of applications in laser science like widely tunable sources employing on-chip tuning \cite{Hua2010,Qin2009,Turcinkova2013}, external cavity \cite{xu2007,hugi2009,riedi2013}, or ultrashort pulse lasers \cite{Keller2003,barbieri2011}.
Broad gain is especially interesting  when combined with locking techniques which enable the access to comb operation. Frequency  combs act as rulers in the frequency domain and are realized starting either from a short-pulse mode-locked  laser \cite{udem_optical_2002_}, or via non-linear processes \cite{DelHayeNature2007,hugi2012}. Combs have been demonstrated in the visible \cite{diddams2010evolving}, Mid-Infrared (Mid-IR) \cite{schliesser2012mid,hugi2012,keilmann2004time}, and Terahertz (THz) \cite{Yasui:JSTQE:2011,Burghoff:NATPHOT:2014} regions of the electromagnetic spectrum. The laser emission from a comb can be stabilized and locked in frequency to highly  stable microwave oscillators and  proficiently used in metrology and high-precision spectroscopy \cite{Holzwarth2000,udem_optical_2002_,Yasui2006,Bernhardt2010}. The mostly used and  also most efficient way to stabilise the offset frequency of a frequency comb is based on the "self-referencing" \(f-2f\) method \cite{diddams2000}, which requires an 
octave 
spanning laser emission. It allows also to determine 
the carrier 
frequency of the comb. Achieving an octave-spanning spectrum is therefore a milestone for any broadband laser.
Until now, octave 
spanning laser emission has been obtained by means of non-linear optics i.e. broadening the laser spectra in a suitable medium \cite{Wadsworth:JOSAB:2002,Bellini:OL:2000}, by using parametric frequency conversion in ultra-high Q factor microresonators \cite{Delhaye:PRL:11} or by directly integrating self phase modulation into Ti:Sapphire oscillators \cite{Ell:OL:2001,Fortier:OL:2003}. An octave spanning semiconductor injection laser, as presented in this work, represents a new attractive opportunity towards compact, on-chip frequency combs. 
The semiconductor quantum cascade laser (QCL) is especially  suitable for direct comb operation without additional locking mechanism required \cite{hugi2012, Burghoff:NATPHOT:2014}. The comb formation is driven by ultrafast nonlinearities (Four-wave mixing) in the active region itself \cite{Khurgin2014}. At THz frequencies the comb operation is additionally favoured by the longer lifetimes for QCLs at those frequencies \cite{Amanti:NJP:09,Scalari:LPR:09}.\\
Here we present a semiconductor injection laser which  operates in the THz range with an emission spanning more than one octave , from 1.64 THz to 3.35 THz  (from 89.5 $\mu$m to 183 $\mu$m in wavelength). The presented devices are monolithic and have typical dimensions of a few millimetre in length and some hundreds of micrometer in width. The laser emission presents no spectral holes all across the 1.71 THz-wide emission bandwidth. At lower injection currents the laser features comb operation with a spectral bandwidth up 624 GHz and a beatnote linewidth of 980 Hz limited by the jitter of the free-running beatnote.\\
\begin{figure}[b]
  \centering
  \includegraphics[width=\textwidth]{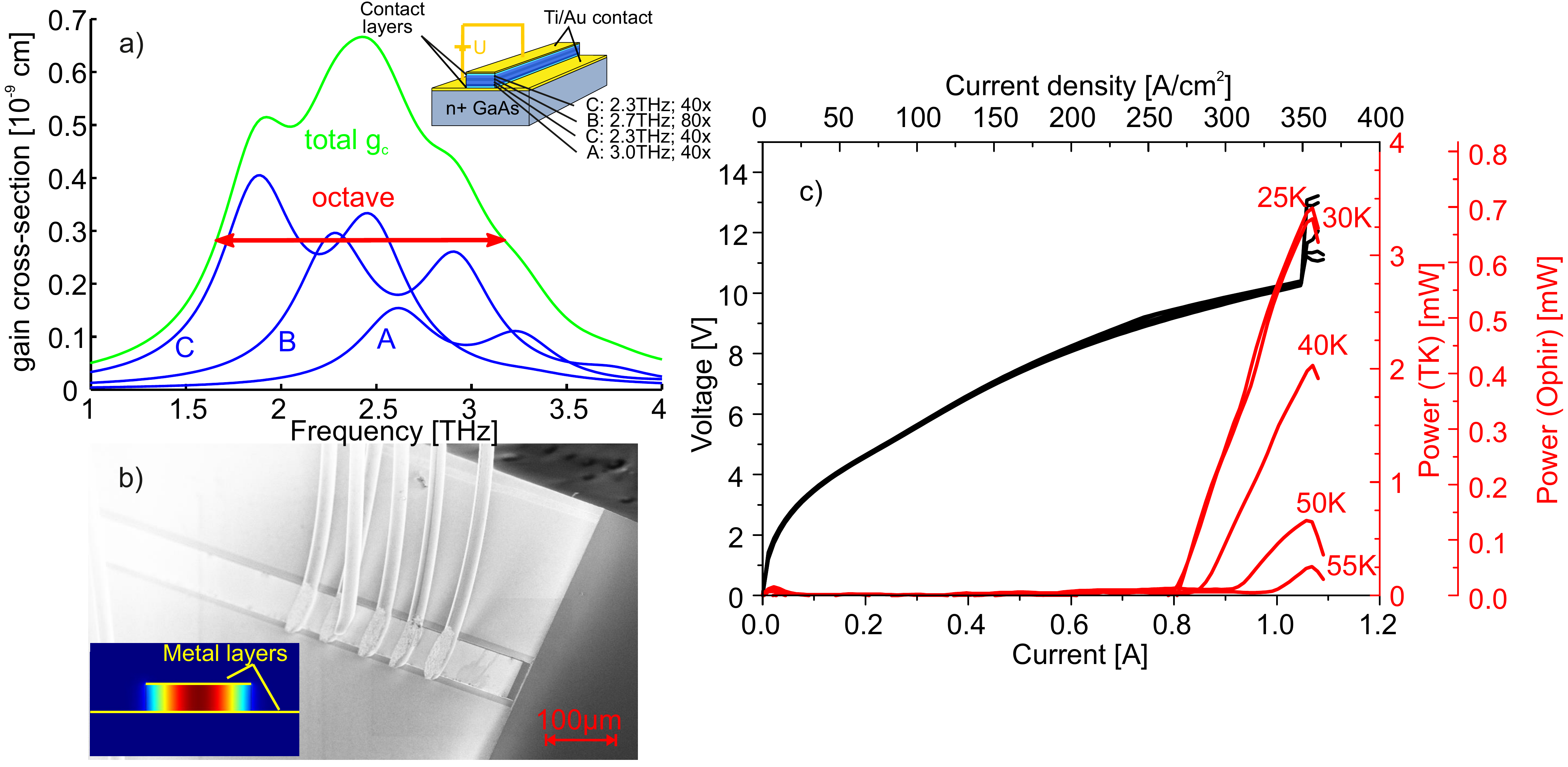}
  \caption{Laser characteristics: a) Calculated \(g_c\): blue curves for the individual designs; green curve for the total active region. Inset: Arrangement of the different active region designs in the laser. b) SEM picture of a processed 50 \(\mu\)m wide dry-etched laser. Inset: Electrical field distribution in a metal-metal waveguide. c) LIV characteristics in CW at different temperatures. The first power axis is normalized to a measurement with a broad area THz absolute powermeter (TK instruments, aperture 55 x 40 mm\(^2\)), the second axis is from an Ophir THz absolute powermeter with smaller chip (aperture diameter 12 mm).}
  \label{fig:sim_sem_spec}
\end{figure}

\subsection*{Results}

\begin{figure}[b]
  \centering
  \includegraphics[width=\textwidth]{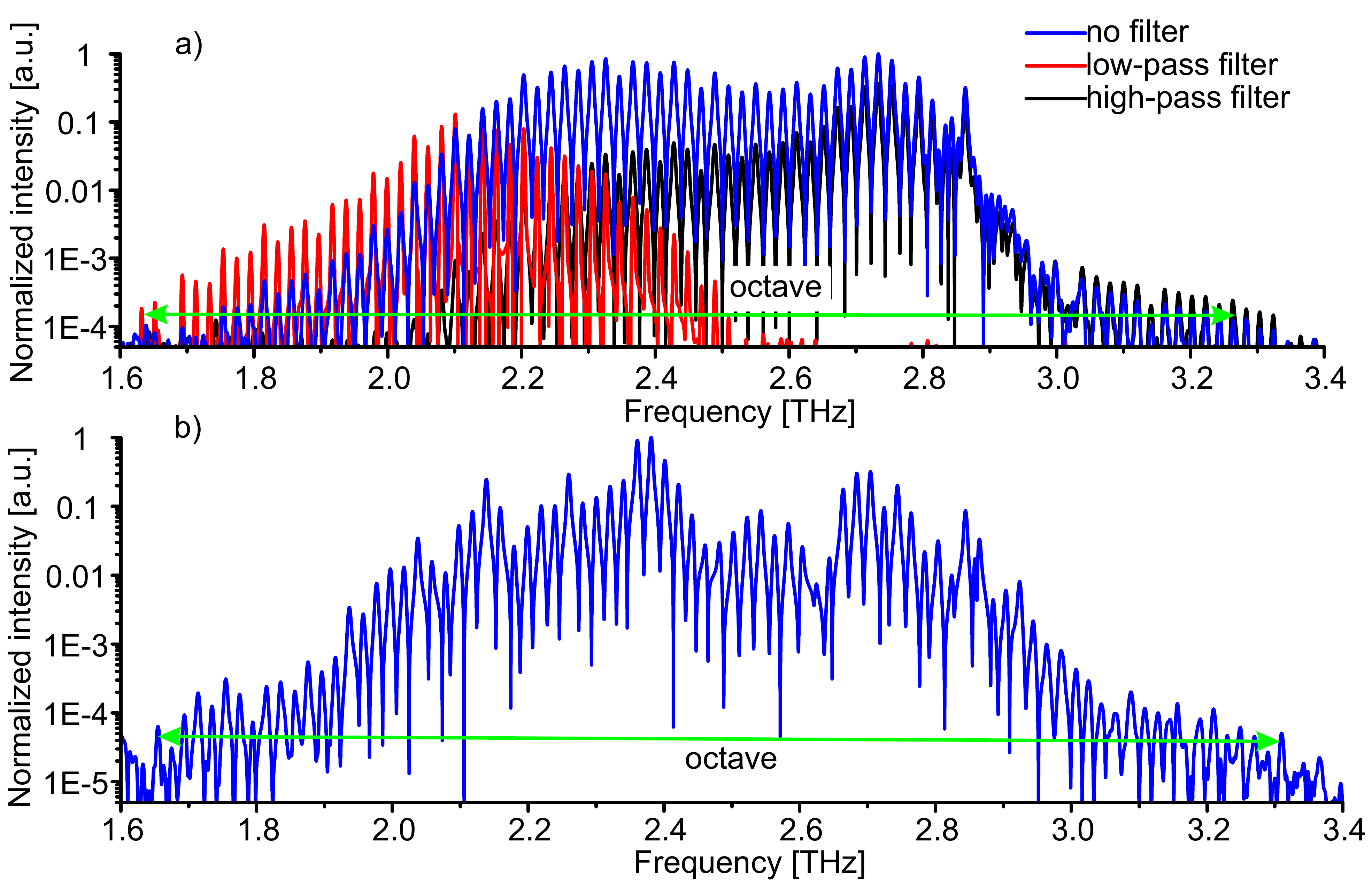}
\caption{Spectral performance: a) Laser spectrum in CW at 25 Kelvin measured with high-pass filter (black), low-pass filter (red), without filter (blue) for a 2 mm x 150 \(\mathrm{\mu m}\) wet-etched laser ridge. c) Octave spanning spectrum of a dry-etched 2 mm x 50 \(\mu\)m laser in CW at 18 Kelvin.}
\label{fig:spec_liv_cw}
\end{figure}
The ultra-broad gain bandwidth is achieved by fully exploiting quantum engineering of intersubband transitions, integrating in the same laser ridge (resonator) different designs of a quantum cascade structure \cite{Faist:Sci:94:553}, tailored at different frequencies. First demonstrated  in mid-IR quantum cascade lasers (QCL) \cite{gmachl2002}, this heterogeneous cascade concept has also been successfully implemented in THz QCLs \cite{Freeman:08,freeman2010,khanna2009} using up to three different designs within one THz QCL \cite{turcinkova2011}. In the present work, we use three different active regions centred at 2.9 THz, 2.6 THz and 2.3 THz that are stacked together, filling the core of a broadband, cutoff-free double metal resonator \cite{Scalari:LPR:09}. 
In order to obtain uniform power across the entire lasing region, special care has been taken in the design of the active regions and the resulting gain profile. We employed a simple model based on the calculation of the spectral gain cross-section \cite{Faist:Book1:2013} g$_{i}=\frac{2\pi e^2 z_i^2}{\epsilon_0 n_{ref} L_i \lambda_i }\frac{\gamma}{(E_i-\hbar\omega)^2+\gamma^2}$  for each substack i \cite{gaincross} (blue curves in Figure \ref{fig:sim_sem_spec}a).  We simulate the complete gain profile of the structure weighting the contribution of each design with the respective overlap factor (the layer sequences for the 3 active regions are reported in Methods). This factor can be simply expressed by  $\Gamma_i=\frac{N_i L_i}{\Sigma_{i=1}^3 N_i L_i}\simeq\frac{N_i}{N_{tot}}$ because in a double metal waveguide the electric field intensity is basically the same throughout the cross-section (see inset of Fig.\ref{fig:sim_sem_spec}b) and the period length for the three active regions is almost identical ($L_
1=L_2=65.6$ nm and $L_3=65.7$ nm). For the total 
spectral gain cross section we obtain g$_{tot}=\Sigma_{i=1}^3\frac{N_i}{N_{tot}}
g_{i}$ which is plotted  
in 
Fig.\ref{fig:sim_sem_spec}a. For broadband operation the most critical aspect in the active region design is that the different stacks have the same maximum current $J_{max}^i=\frac{n_s e}{2 \tau_{up}^i}$ where $n_s$ is the sheet carrier density for the i-th stack and $\tau_{up}^i$ is the respective upper state lifetime.  From band structure calculation and lifetime evaluation including optical phonon scattering we obtain that, at low temperatures, the different lifetimes are all equal within 10$\%$ to the same value $\tau=6$ ps for a lattice temperature T=50 K. This allows  to dope each stack in the same way in order to obtain an n$_s=3.1 \times 10^{10}$ cm$^{-2}$. 
This value of the doping density was  optimised together with the number of periods with respect to our previous devices \cite{turcinkova2011}, in order to reduce the dissipated electrical power and favour continuous wave operation. We employ 40 repetitions of the active region centred at 2.9 THz and 80 repetitions for each of the two lower frequency regions, resulting in a total of N=200 periods and a thickness of the waveguide core  of 13.12 $\mu$m. \\
A first set of lasers was fabricated using wet-etching techniques to define the laser ridges. The second set was processed with dry-etching techniques. The scanning electron microscope (SEM) picture of a 50 \(\mu\)m wide dry-etched laser can be seen in Fig. \ref{fig:sim_sem_spec}b. Dry-etching allows to produce narrow laser ridges without losing spectral bandwidth. This is specially interesting for continuous wave operation (CW) of longer laser cavities. 
Figure \ref{fig:sim_sem_spec}c shows light-current and current-voltage characteristics for a 2 mm x 150 \(\mathrm{\mu m}\) wet-etched laser in continuous wave operation at different temperatures. A maximum power of 3.4 mW in CW at 25 Kelvin is  achieved. The lasing threshold is reached for a current density of \(\mathrm{260 A/cm^2}\). \\
Spectral measurements were carried out with an under-vacuum commercial Fourier transform infrared spectrometer (FTIR) with a resolution of 0.075 cm$^{-1}$. 
A typical spectrum from a  2 mm x 150 \(\mathrm{\mu m}\) wet-etched laser ridge is reported in Fig.\ref{fig:spec_liv_cw}a. In order to rule out any possible spectral artefact coming from non linearities in the measurement setup, we additionally conducted the spectral measurements employing  high and low-pass filters. Like this the strong signal present at the centre of the bandwidth is highly attenuated and the dynamic range of the detector can be fully exploited for the measurement of the weaker modes at the extremes of the bandwidth of the laser emission. We employed a commercial  low-pass multi-mesh filter (QMC instruments: Standard 66 \(\mathrm{cm^{-1}}\) low-pass filter) with a a cutoff at 2 THz and a home made, high-pass mesh filter (0.1 mm thick layer of molybdenum with 60 \(\mu\)m holes) with a cutoff at 2.8 THz. 
As shown in Figure \ref{fig:spec_liv_cw}a the signal-to-noise at both band extremes increases. The fact that the weak modes are still present proves that they are indeed laser signal coming from the laser cavity. \\
The lasing region extends from 1.64 THz to 3.35 THz, covering more than one octave up to a temperature of 30 K. The mode intensity is remarkably well distributed and we can count a total of 84 modes above lasing threshold. The broadband emission is present up to 40 K where the bandwidth is still 1.53 THz. 
Octave-spanning lasing in CW was observed on several laser ridges realised with both dry-etching and wet-etching techniques. An octave spanning spectrum for a dry-etched 2 mm x 50 \(\mu\)m laser is shown in Figure \ref{fig:spec_liv_cw}b.  \\
To study the effects of the laser dimensions on the spectral performance, lasers with different lengths and widths were measured. There is only a minor influence of the cavity length on the spectral bandwidth, as shown in the supplementary figure 1, where we compared spectra of lasers with lengths of 1 mm, 2 mm, 2.2 mm, 3 mm, and 4 mm operating in pulsed mode. Due to the different round-trip frequencies the mode-spacing of the Fabry-P\'erot modes for the individual lasers is different while the spectral range stays almost identical. \\
\begin{figure}[b]
  \centering
  \includegraphics[width=\textwidth]{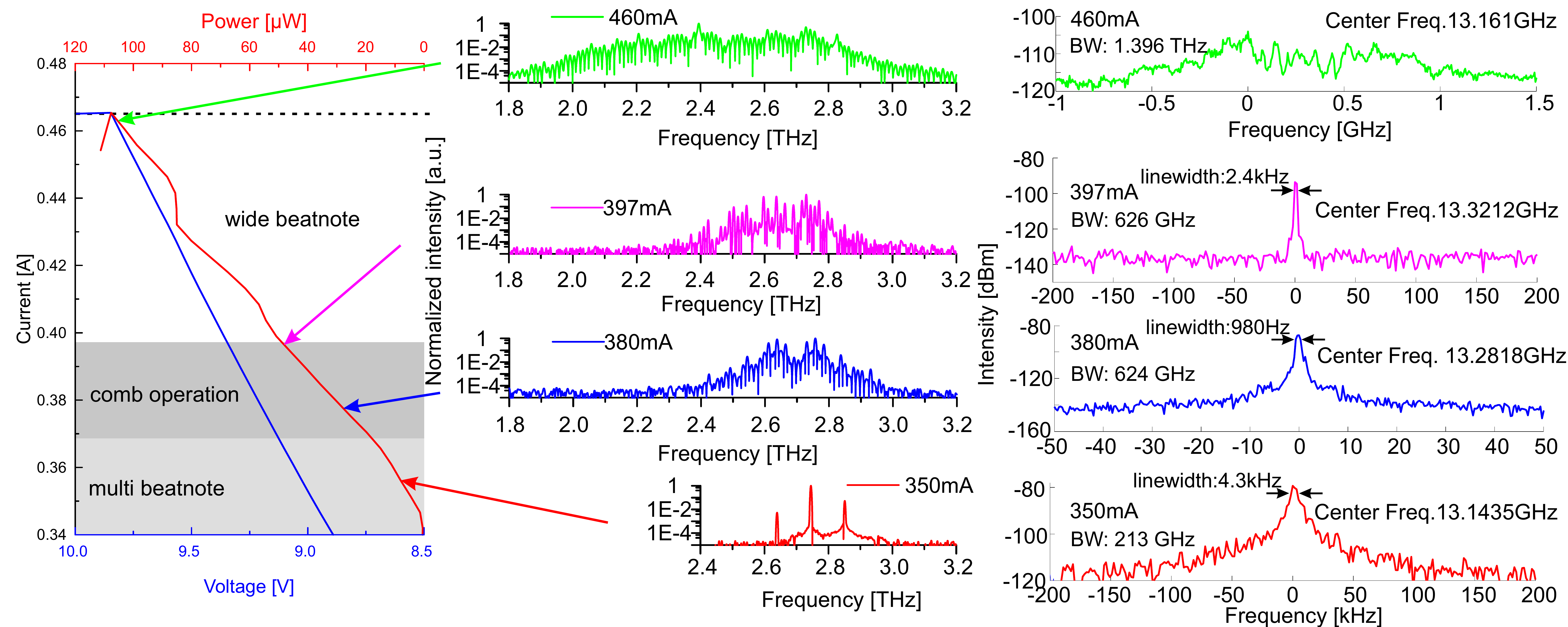}
\caption{Beatnote analysis: Spectral evolution along the LIV for a 3 mm x 50 \(\mu\)m dry-etched laser. Left: the corresponding LIV range where the shaded area indicates the comb region. In the light grey area subcombs are observed. Middle: Spectra for different currents (for a current of 460 mA we have 105 modes above threshold). Right: Corresponding electrical beatnote measured with an antenna.}
\label{fig:beatnote}
\end{figure}
In order to characterize  the spectral emission and its coherence, we performed beatnote measurements at different points of the L-I curve. In Figure \ref{fig:beatnote} we present electrical beatnote measurements (more details in the Methods section) with the corresponding spectral emission in the THz domain for a 3 mm x 50 \(\mu\)m laser. The laser features a region where a collapse of the beatnote linewidth is observed. As already shown experimentally in Refs.\cite{hugi2012,Burghoff:NATPHOT:2014} and discussed theoretically in Ref. \cite{Khurgin2014} the beatnote collapse for a quantum cascade laser is  a clear indication that the laser is acting as a frequency comb. The nonlinearities due to the gain medium itself favour a four-wave-mixing process that  drives the comb formation. In addition, our computations using the model of Ref.\cite{Khurgin2014} indicates that comb operation is slightly favoured by the relatively low upper state lifetime (\(\omega\tau=0.5\) for our 3 mm long device, where \(\omega\) is the roundtrip angular frequency). The beatnote for a CW free-running, non-stabilized laser displays a width of 940 Hz until an injected current of 397 mA, which corresponds to a spectral bandwidth of 628 GHz with 47 modes above threshold. The linewidth is 
limited by the jittering of the beatnote since the laser is unstabilized. This compares well with the results reported in Ref. \cite{Burghoff:NATPHOT:2014} where a stabilized
comb spanning less than 500 GHz with spectral holes was reported with beatnote widths of 1.53 KHz. 
\begin{figure}[tb]
  \centering
  \includegraphics[width=0.6\textwidth]{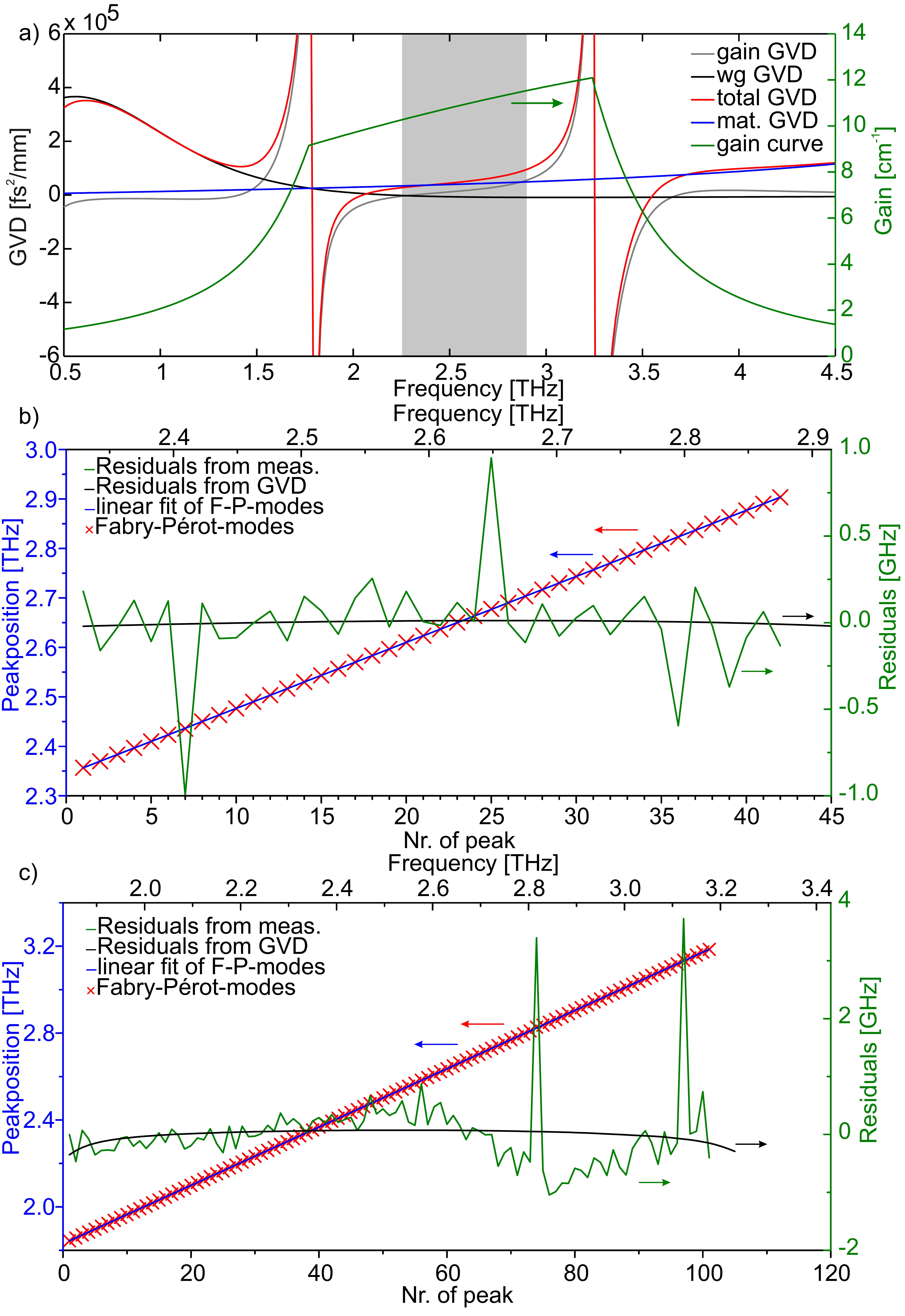}
  \caption{Dispersion analysis: a) GVD simulations including contributions from the material (GaAs), waveguide and gain. The shaded part indicates the spectral range of the comb-regime. (b,c) Residuals to a linear fit of the measured peak positions (c,e) for a 3 mm x 50 \(\mu\)m laser at 397mA and 460mA in cw operation compared to the same quantity calculated theoretically from the GVD showing the effect of dispersion in the lasing spectra.}
  \label{fig:residuals}
\end{figure}
For currents just above threshold (see Fig.\ref{fig:beatnote} ) we observe several narrow beatnotes spaced by hundreds of MHz from each other. Most probably this can be attributed to  optical feedback effects. Similar effects were also reported for microresonator combs \cite{Delhaye:PRL:11}.
When not facing the FTIR (and thus reducing optical feedback) the comb regime expands up to 422 mA (0.92 \(\mathrm{I_{NDR}}\)). For higher currents the beatnote progressively broadens as the emission bandwidth covers a wider spectral region. This is the effect of the large dispersion introduced by both the GaAs, due to the proximity of the Reststrahlenband, and the contribution of the active region.  As already shown in Ref.\cite{Burghoff:NATPHOT:2014} dispersion management is a crucial point to address, when aiming at comb operation.
The comb behaviour can be understood when carefully calculating the total group velocity dispersion (GVD) in the laser according to Ref.\cite{hugi2012}, taking into account the material dispersion of GaAs together with the dispersion introduced by the waveguide and the one from the laser gain (see Methods for details). As shown in Fig. \ref{fig:residuals}a the GVD from the waveguide partially compensates for the material GVD of GaAs. Together with the flat gain introducing an almost zero GVD at the centre laser frequencies a region with low and flat GVD is achieved (red curve in Fig. \ref{fig:residuals}a). The maximal measured spectral emission for the comb operation is indicated by the shaded area.
As long as the laser is only operating in this region, i.e. at low currents, it acts as a comb. As soon as the bandwidth broadens too much, dispersion is coming into play and the comb-regime is left as indicated by the broadened beatnote.

\subsection*{Discussion}

A quantitative assessment  of dispersion can be given by examining the mode-spacing of the Fabry-P\'erot modes. For zero dispersion the mode-spacing is constant over the entire frequency range. The laser then acts as a frequency comb \cite{hugi2012,Burghoff:NATPHOT:2014}. 
 Group velocity dispersion (GVD) causes a change in the mode-spacing and therefore one can deduce the GVD in the laser cavity by measuring the mode-spacing. The limitation for measuring the mode-spacing in our laser is the resolution of the FTIR (\(\mathrm{0.075cm^{-1}}\), 2.25 GHz). Within this resolution no change in the mode-spacing is observed in the laser. To overcome this limitation one can use a zero-padding to perform the Fourier-transform of the FTIR interferogram. The zero-padding will smooth the spectrum by generating more points in the spectral curve and therefore give a preciser information of the individual peak positions without changing the resolution of the measurement \cite{griffiths2007}. By taking the difference between the individual modes one gets the mode-spacing displayed in figure \ref{fig:residuals}b and \ref{fig:residuals}d. 
 The shape of the mode-spacing curve is slightly changing when the laser gets into the broad beatnote regime. It therefore indicates the onset of dispersion in the laser.   
 A more visual method to indicate the dispersion is to plot the peak positions as a function of the peak number (lowest frequency peak is no.1) and then interpolate this curve with a linear fit. The residuals between the linear fit and the peak positions (green curve in Fig. \ref{fig:residuals}c and \ref{fig:residuals}e) are therefore again an indicator of the dispersion. For no dispersion the residual curve oscillates around zero. As soon as the broad beatnote regime is reached (i.e. onset of dispersion), there is a trend visible in the curve indicating the frequency dependence of the mode-spacing.
 Only in the regime with a broad electrical beatnote such trends are visible in the residual plots. We are therefore able to check for any spectrum whether the laser is in a zero dispersion regime or not. \\
 From our simulation of the GVD we are also able to predict the modespacing and the residual curves for our laser. A value for the group velocity is fixed and from the GVD one gets the deviations of this value. With that and knowing the length of the laser cavity (3 mm), we calculated the mode-spacing (red curves in Fig.\ref{fig:residuals}b and d) as well as the residual curves (black curves in Fig.\ref{fig:residuals}c and Fig.\ref{fig:residuals}d) theoretically. The simulation agrees with the flat residual curve in Fig. \ref{fig:residuals}c and also indicates a slight curvature at 460 mA (Fig.\ref{fig:residuals}e). The local minimum in Fig.\ref{fig:residuals}e is not predicted by our model. It might result from higher order cavity modes invisible in the spectra due to the resolution of the FTIR which affect the lower order modes by pulling effects. \\ 
 The next steps of our research will be devoted to the implementation of dispersion compensation elements \cite{Burghoff:NATPHOT:2014} to demonstrate comb operation with sub-kHz beatnote linewidth on the entire octave spanned by our lasers.\\
To summarize, we presented an octave-spanning semiconductor laser emitting in the THz range with no spectral holes both in pulsed and CW operation. To the best of our knowledge this is the first octave spanning semiconductor laser. In addition, the laser features a comb regime with sub-kHz beatnote and a corresponding spectral emission of more than 600 GHz bandwidth. 

\subsection*{Methods}

\textbf{Quantum cascade layer sequence and details}.
Sample EV1913 has been grown by Molecular Beam epitaxy on a semi-insulating GaAs substrate in the GaAs/AlGaAs material systems. The layer sequence for the 2.9 THz active region (40 repetitions) is, starting from the injection barrier: ${\bf 5.5}/11.0/{\bf 1.8}/11.5/{\bf 3.8}/9.4/{\bf 4.2}/18.4$. 
The figures in bold face represent the Al$_{0.15}$Ga$_{0.85}$As barrier  and the 18.4 nm GaAs quantum well is homogeneously Si doped $1.7 \times 10^{16}$ cm$^{-3}$. The layer sequence for the 2.6 THz active region (80 repetitions) is, starting from the injection barrier: ${\bf 5.5}/11.3/{\bf 1.8}/11.3/{\bf 3.8}/9.4/{\bf 4.2}/18.4$.  The 18.4 nm GaAs quantum well is homogeneously Si doped $1.7 \times 10^{16}$ cm$^{-3}$. 
The layer sequence for the 2.3 THz active region (80 repetitions) is, starting from the injection barrier: ${\bf 5.5}/12.0/{\bf 1.8}/10.5/{\bf 3.8}/9.4/{\bf 4.2}/18.4$.  The 18.4 nm GaAs quantum well is homogeneously Si doped $1.7 \times 10^{16}$ cm$^{-3}$.

\textbf{Dispersion due to gain profile}. The contribution to the dispersion coming from the broadband gain has been evaluated as follows: at threshold the gain equals the losses so we calculated the frequency dependent losses coming from the waveguide including contact layers (with 2D numerical simulations) then we summed up the frequency dependent contributions due to intersubband absorption in the active region and also the frequency dependent mirror losses evaluated with a 3D numerical calculation. Everything has been summed and then integrated in a gain profile function which includes lorentzian-like tails coming from unclamped gain of true active regions. By applying Kramers-Kroning relations we proceeded to calculate the GVD in the same way as in Ref. \cite{hugi2012}.

\textbf{Beatnote measurement}.
The CW laser was left free running without any active stabilisation. Care was taken to provide stable power supply employing a commercial, highly stabilised source (Wavelength electronics Model QCL1000) followed by a low pass filter. 
The beatnote was measured by picking up the RF signal irradiated  outside of the cryostat with an home-made folded dipole antenna, subsequently  attached to a microwave spectrum analyser (Rohde \& Schwarz FSU50). In the narrow beatnote regime the spectrum analyser was set with a resolution bandwidth (RBW) of 1 kHz (100 Hz at 380 mA) and a video bandwidth of 3 kHz (500 kHz at 380 mA). The traces were recorded in a single sweep mode. The signal was amplified by 35 dB using an internal preamplifier. For the beatnote measurement at 460 mA an average of 100 sweeps was recorded with RBW 200 kHz and VBW 500 kHz also using the internal preamplifier.

\subsection*{Acknowledgement}
The presented work is part of the EU research project TERACOMB (Call identifier FP7-ICT-2011-C, Project No.296500). The funding is gratefully acknowledged. We acknowledge discussions with G. Villares and S. Barbieri and we would like to thank C. Bonzon for his help with the FE simulations.


\begin{thebibliography}{37}
\bibitem{gaincross}
\bibinfo{note}{In the gain cross section the symbology is as follows: n=3.6 is
  the refractive index, , $\lambda_i$ the wavelength for each relevant
  transition in each stack i, $\gamma=1$ meV is the level broadening, z$_{i}$
  is the dipole matrix elements for the two states in each stack which
  contribute to the gain, $e$ is the electron charge and E$_i$ is the energy of
  the relevant transitions in each stack i}.
\end{thebibliography}
\end{document}